\documentclass[journal]{IEEEtran}
\usepackage{float}
\usepackage{graphicx, subcaption, caption}
\usepackage{amsmath}
\usepackage{cite}

\usepackage{lipsum} % For some dummy text
\usepackage{filecontents} % To make the

\begin{document}
\bstctlcite{IEEEexample:BSTcontrol}

\title{Dual Polarization Full-Field Signal Waveform Reconstruction Using Intensity Only Measurements for Coherent Communications}

\author{Haoshuo Chen,
    Nicolas K. Fontaine,
    Joan M. Gene,
    Roland Ryf,
    David T. Neilson,
    Gregory Raybon

\thanks{H. Chen, N.~K. Fontaine, J.~M. Gené, R. Ryf, D.~T.~Neilson, G. Raybon are with Nokia Bell Labs, 791 Holmdel Rd., Holmdel, NJ 07733, USA haoshuo.chen@nokia-bell-labs.com.}

\thanks{J.~M. Gene is also with Universitat Politècnica de Catalunya, Jordi Girona 1-3, Barcelona, 08034 Spain}

\markboth{Journal of \LaTeX\ Class Files,~Vol.~14, No.~8, August~2015}%
%\specialpapernotice{(Top-Scored Paper)}
}

\maketitle

\begin{abstract}

Conventional optical coherent receivers capture the full electrical field, including amplitude and phase, of a signal waveform by measuring its interference against a stable continuous-wave local oscillator (LO). 
In optical coherent communications, powerful digital signal processing (DSP) techniques operating on the full electrical field can effectively undo transmission impairments such as chromatic dispersion (CD), and polarization mode dispersion (PMD).
Simpler direct detection techniques do not have access to the full electrical field and therefore lack the ability to compensate for these impairments.
We present a full-field measurement technique using only direct detection that does not require any beating with a strong carrier LO.
Rather, phase retrieval algorithms based on alternating projections that makes use of dispersive elements are discussed, allowing to recover the optical phase from intensity-only measurements.
In this demonstration, the phase retrieval algorithm is a modified Gerchberg–Saxton (GS) algorithm that achieves a simulated optical signal-to-noise ratio (OSNR) penalty of less than 4~dB compared to theory at a bit-error rate of 2$\times 10^{-2}$. 
Based on the proposed phase retrieval scheme, we experimentally demonstrate  signal detection 
and subsequent standard 2$\times$2 multiple-input-multiple-output (MIMO) equalization 
of a polarization-multiplexed 30-Gbaud QPSK transmitted over a 520-km standard single-mode fiber (SMF) span.
\end{abstract}


\IEEEpeerreviewmaketitle

\section{Introduction}

Full field digitization of optical waveforms by using coherent detection enables the use of powerful digital signal processing (DSP) techniques to decode complex signal constellations~\cite{kikuchi_fundamentals_2016,essiambre_capacity_2010,ip_coherent_2008} such as 16-QAM (Quadrature Amplitude Modulation) and probabilistic constellation shaping~\cite{buchali_rate_2016} and to undo transmission impairments such as chromatic dispersion~\cite{savory_electronic_2007} and polarization mode dispersion.
Detecting the full field of both received polarizations enables multiple-input multiple-output (MIMO) processing to undo polarization mode dispersion effects. 
Coherent receivers are considered costly due to the requirement of a local oscillator (LO) laser and multiple balanced photodetectors which in cost sensitive applications are undesirable.
Unfortunately, simpler direct detection schemes that only have access to the optical intensity such as on-off keying or PAM4 cannot compete in both reach and capacity with coherent detection.
As a result, many complex direct detection receivers have been proposed including the Kramers–Kronig and Stokes-space receivers~\cite{antonelli_kramerskronig_2018,chen_kramerskronig_2018,zhu_direct_2018,che_direct_2018} to extend the reach and increase the capacity of direct detection systems.
Although these solutions use different receiver/transmitter architectures and various DSP algorithms, those that can measure the full field of the modulated signal and perform  equalization require some form of single-side band heterodyne detection using a continuous wave carrier, which is normally added at the transmitter before transmission~\cite{zhu_direct_2018,che_direct_2018,chen_kramerskronig_2018}.

Self-coherent detection schemes~\cite{liu_digital_2008,nazarathy_self-coherent_2008} can recover differentially coded complex signals while eliminating the usage of the optical LO.
However, self-coherent implementations have yet to demonstrate full compensation of impairments such as CD and polarization modal dispersion (PMD) and are constrained to specific baud rates~\cite{ip_coherent_2008,kikuchi_chromatic_2010}.
Recently, compressed sensing techniques that exploit signal sparsity in combination with phase retrieval methods~\cite{moravec_compressive_2007,mukherjee_iterative_2012,hayakawa_reconstruction_2018,arik_low-complexity_2018} have been proposed. %to realize sub-Nyquist rate signal sampling and recovery
However, they make the received signal appear sparse by using numerous linear projections of the signal, which increases the receiver complexity as it requires a large number of parallel receivers ~\cite{sefler_demonstration_2018,yoshida_coherent_2019,scofield_speckle-based_2019}.  
Using dispersion as a projection does not rely on sparsity.
Dispersive propagation can convert phase modulation to amplitude modulation, which has been widely applied in various fields, like for example in the characterization of ultrashort optical pulses~\cite{dorrer_characterization_2005,cuadrado-laborde_phase_2014}.
The use of dispersion as an alternative projection method to recover temporal phase has been demonstarted in~\cite{matsumoto_phase_2018, blech_enhanced_2018}.
However, an optical carrier is still needed and only simulation results were presented.

Here, we show a direct detection technique that can measure most complex-valued waveforms interesting for telecommunications such as Nyquist-shaped Quadrature phase shift keying (QPSK) and QAM signals without relying on any optical carriers~\cite{chen_full-field_2019}.
It uses phase-retrieval concepts borrowed from X-ray crystallography, image processing and astronomy~\cite{burvall_phase_2011,shechtman_phase_2015,jaganathan_phase_2015} to reconstruct complex-valued signals from two temporal intensity measurements taken before and after a dispersive element~\cite{cuadrado-laborde_phase_2014}.
It uses the same number of receivers (single-ended photo-diodes) as coherent receivers (balanced photo-diodes) but eliminates the 90-degree optical hybrids and the LO.
Performing the phase recovery on both received polarizations or multiple received spatial modes allows the phase-retrieval receiver to act analogously to a standard coherent receiver supporting both polarization-division multiplexing and space-division multiplexing~\cite{van_weerdenburg_138-tb/s_2018} using only direct detection.
As a demonstration, we measure a polarization-multiplexed 30-Gbaud QPSK signal after 520-km single-mode fiber (SMF) transmission with a line rate of 120-Gbit/s, which has accumulated strong polarization scrambling (PMD) and CD.

\section{Single Polarization Phase Retrieval and Algorithm}
\begin{figure}[!t]%1
\centerline{\includegraphics[width=2.8in]{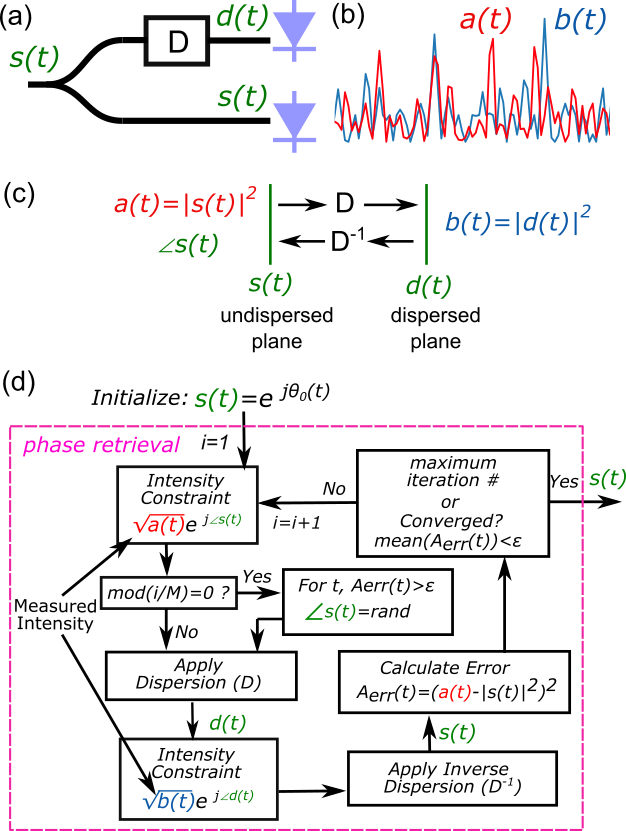}}
\caption{~(a) Illustration of the phase-retrieval receiver for detecting a single-polarization complex-valued signal based on direct detection (D: dispersive element),
(b)~example intensity of QPSK symbols measured after photodetectors with and without experiencing optical dispersion,
(c)~schematic of undispersed and dispersed signals before and after dispersion, related by a paired functions: $D$ and $D^{-1}$,
and (d)~modified Gerchberg–Saxton (GS) algorithm diagram with local-minima escape for phase retrieval.
}
\label{FIG:1}
\end{figure}

The receiver uses a phase retrieval algorithm to recover the full complex valued field from two intensity measurements taken before and after a dispersive element.
Figure~\ref{FIG:1}(a) shows the schematic of the phase-retrieval receiver comprising a splitter, dispersive elements, and two photodetectors.
The two intensity measurements denoted as $a(t)$ and $b(t)$ are digitized and shown in Fig.~\ref{FIG:1}(b).
Figure~\ref{FIG:1}(c) shows how the dispersive element, $D$, relates the full-field before dispersion, $s(t)$, and after dispersion, $d(t)$.

A modified Gerchberg–Saxton (GS) algorithm~\cite{gerchberg_practical_1972,fienup_phase_1982} iteratively refines the estimate of the phase of the undispersed waveform, $s(t)$, until its intensity after numerical propagation matches the dispersed measurement $b(t)$ as illustrated in Fig.~\ref{FIG:1}(d).
Intensity measurements $a(t)$ and $b(t)$ are normalized by their maximum value.
The algorithm is initialized as iteration number $i$=1 with complex field $s(t)=e^{j\theta_0(t)}$, where $s(t)$ is a random complex vector.
An improved estimate of the field is constructed from the directly detected intensity $a(t)$ and the phase of $s(t)$ (i.e., $\sqrt{a(t)}e^{j \angle s(t)}$ ).
This field is propagated through a dispersive element which is represented by a quadratic spectral phase filter to produce an estimate of the dispersed field, $d(t)$.
Replacing the intensity of $d(t)$ with the second intensity measurement, $b(t)$, improves the estimate of $d(t)$ and it is propagated through inverse dispersion $D^{-1}$ to produce a new guess of the field, $s'(t)$.
After each iteration, the phase retrieval error, $A_{err}(t)= |a(t)-|s'(t)|^2|^2$ is calculated.

GS algorithms often get trapped into local minima, and finding the global minimum is one of the toughest challenges for phase retrieval algorithm.
To escape from local minima~\cite{fienup_phase-retrieval_1986} a random number is assigned to the phases of the samples with the phase retrieval error $A_{err}(t)$ above $\epsilon$ ($A_{err}(t)>\epsilon$) after every $M$ iterations, where $\epsilon$ is an acceptable error level.
This process repeats until either the iteration number reaches the maximum or the mean of $A_{err}(t)$ is minimized below $\epsilon$.
$\epsilon$ can be pre-determined based on the system delivered optical-signal-noise ratio (OSNR).
% (See Supplement for details).
The algorithm also includes a spectral constraint to force a rectangular spectrum for our Nyquist shaped waveforms and pilot symbols to help with convergence, which are omitted in Fig.~\ref{FIG:1}(b) for the sake of simplicity.
The pilot symbols eliminate the constant phase ambiguity across the entire waveform, and assists the algorithm in converging to the optimum solution, especially in the presence of noise.
CD from the transmission fiber \textit{$h_{CD}$} is compensated prior to applying the pilot symbols.
% (See Supplement for details).
The details of the applied spectral constraint, pilot symbols and \textit{$h_{CD}$} compensation are discussed in Appendix A.

\begin{figure}[!b]%1
    \centerline{\includegraphics[width=3.5 in]{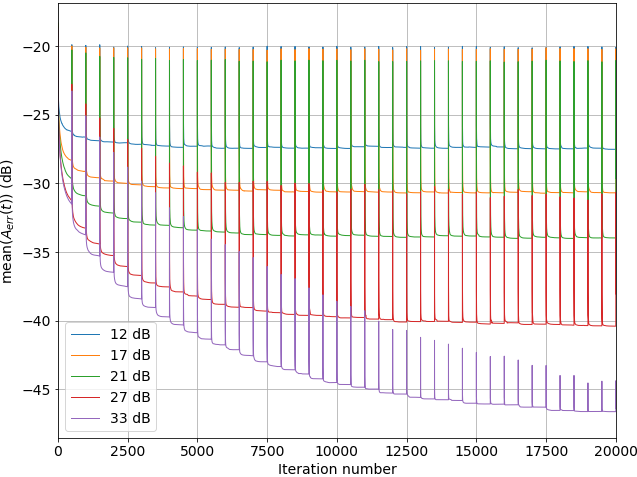}}
    \caption{~Mean $A_{err}(t)$ averaged over $2^{11}$ symbols in dB versus the number of iterations for different OSNRs (local-minima escape is reset after every 500 iterations, which induces the spikes).
    }
    \label{FIG:2a}
\end{figure}

\begin{figure*}[!t]%1
\centerline{\includegraphics[width=6.8in]{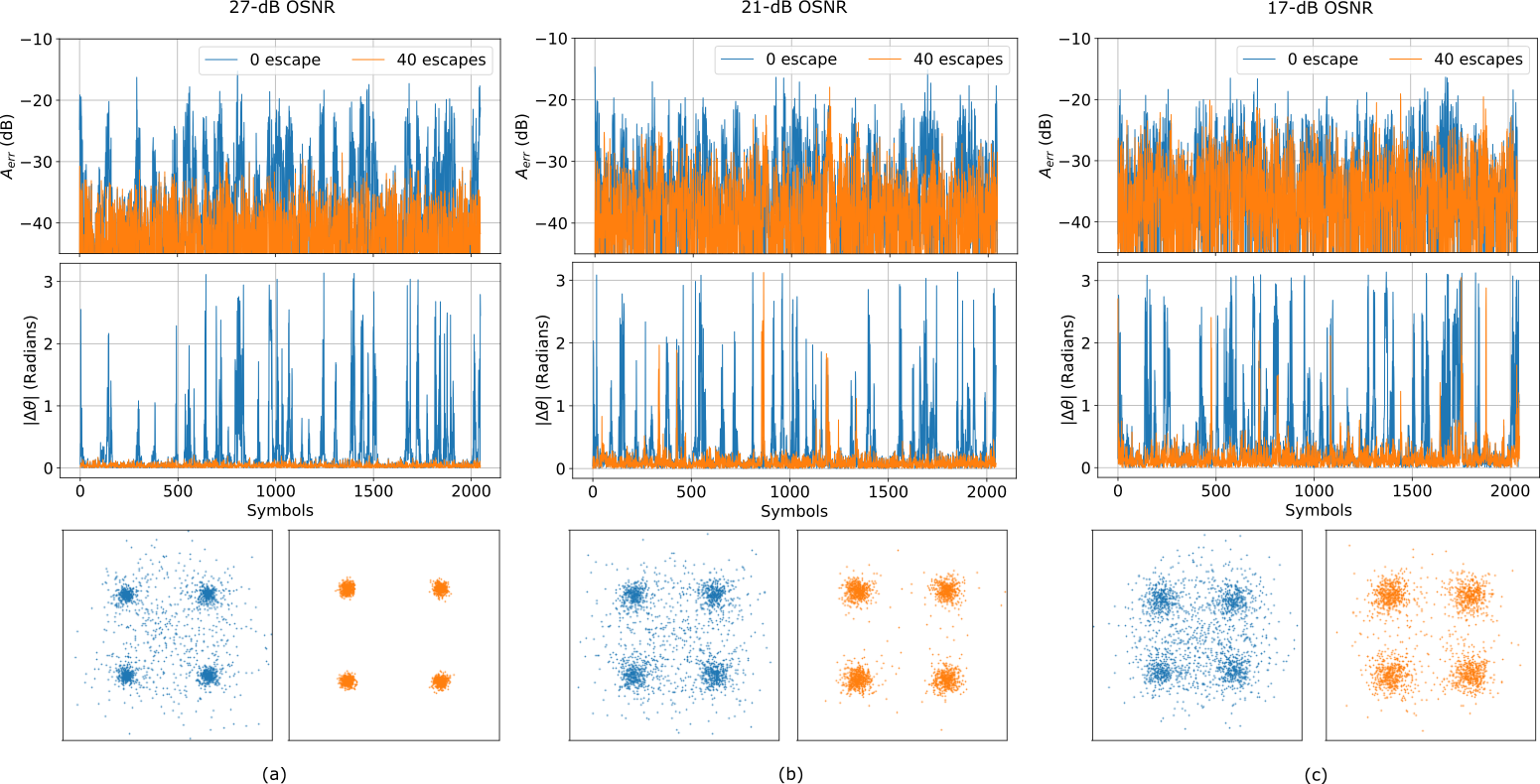}}
\caption{~Simulation results for a single-polarization 30-Gbaud QPSK signal at (a) 27, (b) 21 and (c) 17-dB OSNR after 60-km SMF transmission: 
[upper] $A_{err}$, [middle] absolute phase error $|\Delta\theta|$ for all the symbols and [lower] reconstructed constellations of the QPSK signal after 0 and 40 local-minima escapes.
}
\label{FIG:2}
\end{figure*}

\begin{figure}[!b]%1
    \centerline{\includegraphics[width=3.0 in]{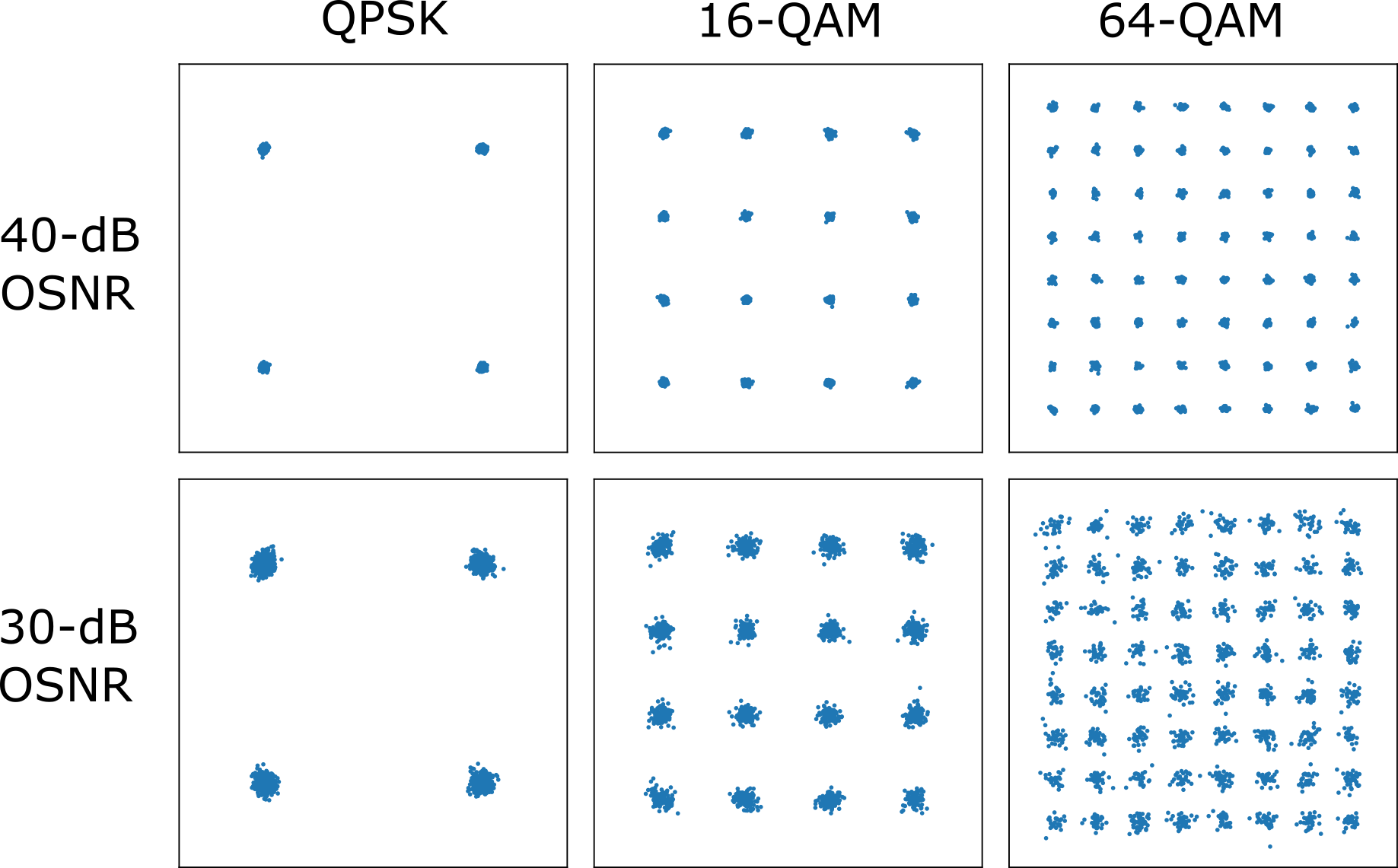}}
    \caption{~Reconstructed constellations for different modulation formats: QPSK (left), 16-QAM (middle) and 64-QAM (right) for 40-dB (upper) and 30-dB (lower) OSNR.
    }
    \label{FIG:2b}
\end{figure}

\begin{figure}[!b]%1
\centerline{\includegraphics[width=3.0in]{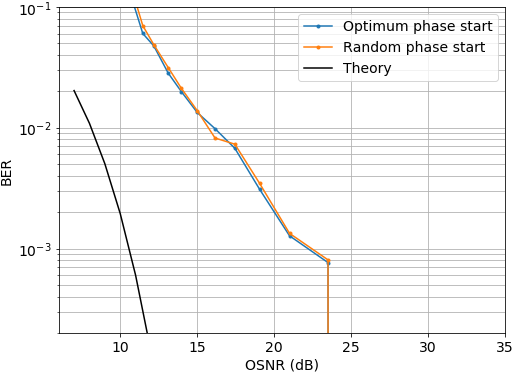}}
\caption{~BER versus OSNR using optimum and random phase start for phase retrieval, respectively.
}
\label{FIG:3a}
\end{figure}

\begin{figure}[!t]%1
\centerline{\includegraphics[width=2.8in]{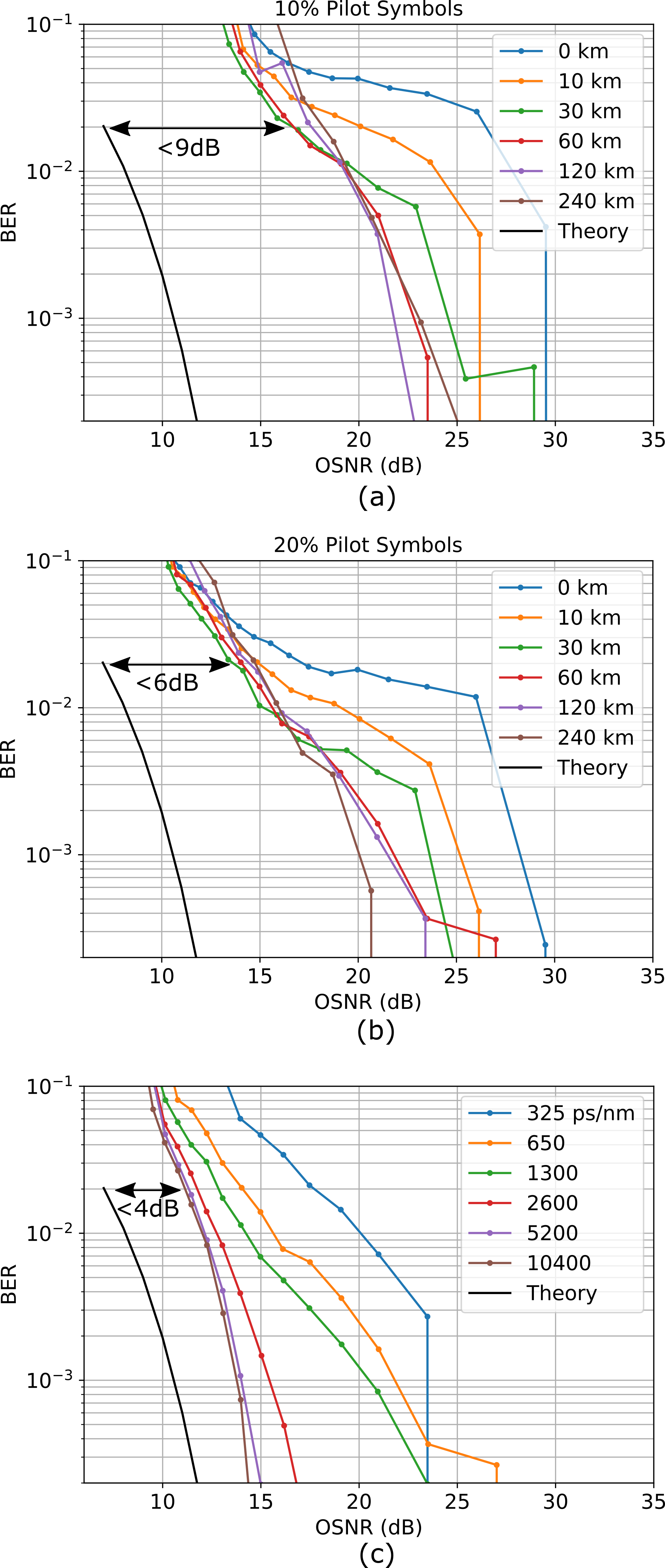}}
\caption{~Simulated BER versus OSNR for a single-polarization 30-Gbaud QPSK signal: after transmission over SMF with various lengths using (a) 10$\%$ and (b) 20$\%$ pilot symbols for phase retrieval,
and (c) for different amount of dispersion applied in phase retrieval after 60-km SMF transmission and 20$\%$ pilot symbols.
}
\label{FIG:3}
\end{figure}

\begin{figure}[!t]%1
    \centerline{\includegraphics[width=3.5in]{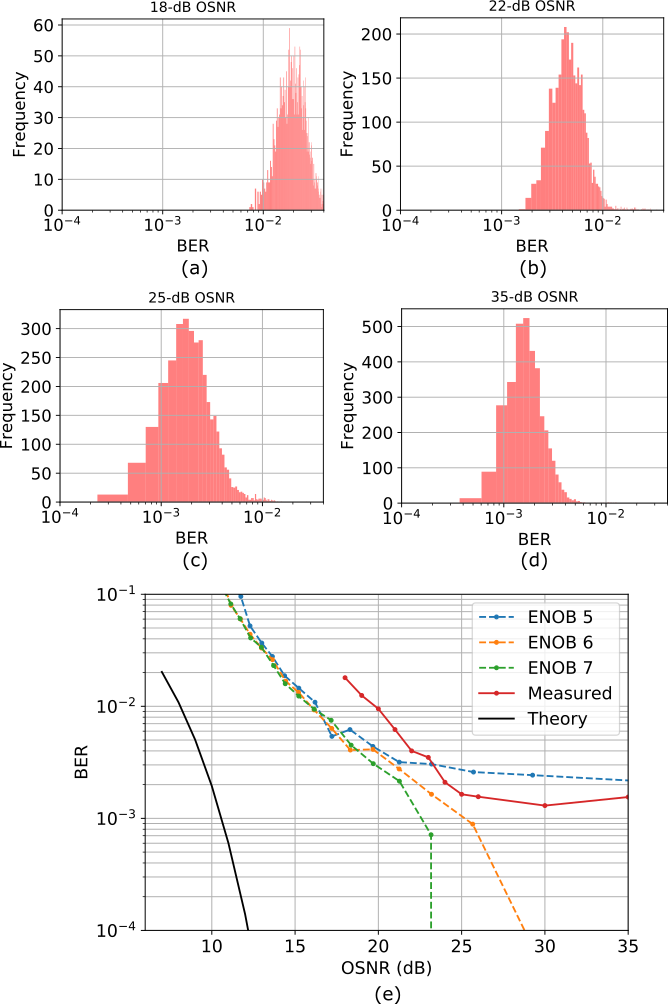}}
    \caption{~Experimental results of a 30-Gbaud QPSK signal after 60-km SMF transmission: (a)-(d) histogram of the measured BERs under different OSNR,
    (e) BER curve versus OSNR for the measurements and simulations with different ENOB.
    }
    \label{FIG:4}
\end{figure}

\section{SINGLE-POLARIZATION SIMULATION RESULTS}

\subsection{Simulations and Numerical Procedure}

We first perform numerical simulations to study the performance of the LO-less phase-retrieval receiver compared to theoretical limits.
Simulation results for a single-polarization Nyquist-shaped 30-Gbaud QPSK signal at different OSNRs after 60-km single-mode fiber (SMF) transmission are presented in Fig.~\ref{FIG:2}.
The OSNR is calculated at 0.1-nm measurement bandwidth.
$2^{11}$ symbols are modeled and are pulse shaped using a raised-cosine filter with a roll-off of 0.1 and sampled at 60 GSamples/s to achieve higher spectral efficiency.
A dispersive element introducing 650~ps/nm CD is applied for phase retrieval, which spreads a 30-Gbaud optical pulse over more than 8 symbols.
Every 500 iterations, phase resets are applied to the symbols with $A_{err}>\epsilon$ to escape the local minima.
$\epsilon$ is set to be $2\times10^{-3}$.
The phases of the pilot symbols (20\% overhead) are fixed to the predetermined values to facilitate convergence.

Figure~\ref{FIG:2a} shows the mean $A_{err}(t)$ averaged over all the symbols in dB versus the number of escapes.
% Below -40-dB mean $A_{err}(t)$ can be achieved after 10 escapes.
The level of the mean $A_{err}(t)$ after convergence is determined by the OSNR so that the mean $A_{err}(t)$ after convergence can be pre-calculated by the measured OSNR and used for determining $\epsilon$, which indicates when the phase retrieval algorithm converges and can be terminated.
% (See Supplement). 
Figure~\ref{FIG:2} shows $A_{err}$ and absolute phase error $|\Delta\theta|$ for all the symbols after 0 and 40 escapes at different OSNRs.
$\Delta\theta$ is the calculated angle difference between $s(t)$ and the signal, which is smaller than 0.1 radians after 40 escapes.
$A_{err}$ and $|\Delta\theta|$ evolution during the entire phase retrieval process is present in Appendix C.
Figure~\ref{FIG:2} also provides the corresponding reconstructed constellations of the QPSK signal.
The proposed phase retrieval algorithm is applicable to other high-order modulation formats.
The recovered constellations of Nyquist-shaped 16-QAM and 64-QAM signals are shown in Fig.~\ref{FIG:2b}.

\subsection{Receiver Performance vs. Optical Signal to Noise Ratio}

It is verified through simulation that the phase retrieval algorithm can converge to the optimum state irrespective to the phase initialization, see Fig.~\ref{FIG:3a}, where Nyquist-shaped 30-Gbaud QPSK signal after 60-km SMF transmission is recovered using 650 ps/nm dispersion.
The optimum phase start condition means that the exact phases of the transmitted symbols are used as the initialized phases in these simulations.
For the rest of the simulations, the initial phase is randomly generated with a uniform distribution over -$\pi$ to $\pi$.
The theoretical curve is calculated based on Shannon's information theory for an additive white Gaussian noise (AWGN) channel assuming only optical noise, which is usually displayed as a performance reference for optical coherent communications.

Figure~\ref{FIG:3}(a) and (b) gives the simulated BER versus OSNR after SMF transmission with different distances.
For transmission less than 25-km, the receiver performs worse than after 25-km propagation because the initial dispersion \textit{$h_{CD}$} provides some additional amplitude randomness which is proven to provide more information to the phase retrieval algorithm and improve convergence~\cite{xie_chromatic_2013}.
Thus, system performance for shorter distance transmission could be improved by sending a pre-dispersed signal~\cite{killey_electronic_2005}.
The OSNR penalty at BER = 2$\times 10^{-2}$ using a 650 ps/nm dispersion is around 6 and 9 dB employing 20\% and 10\% pilot symbols, respectively.

Figure~\ref{FIG:3}(c) shows the BER results versus the dispersion applied in the phase retrieval after 60-km SMF transmission.
Phase recovery accuracy improves with larger dispersion, which mixes more signal symbols together, and the BER converges to within 4~dB to the theoretical limit using 5200~ps/nm of dispersion.
It draws an analogy to multi-symbol self-coherent detection where recovered phase error can be reduced through an averaging effect over multiple symbols~\cite{nazarathy_self-coherent_2008}.
It can also be interpreted that interfering more symbols together reduces the possibility that the phase retrieval algorithm becomes trapped in sub-optimal solutions.

\begin{figure*}[!t]%1
\centerline{\includegraphics[width=6.0in]{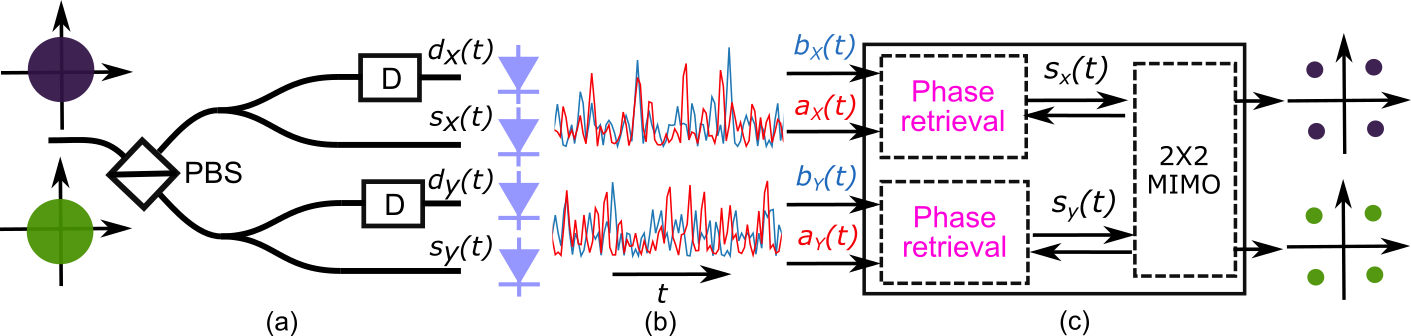}}
\caption{(a) Schematic of the polarization-diversity phase-retrieval receiver for detecting a polarization-diversity complex-valued signal based on direct detection (D: dispersive element, PBS: polarization beam splitter),
    (b) example intensity measured after photodetectors,
    and (c) digital signal processing (DSP) diagram including phase retrieval, equalizer feedback and 2 $\times$ 2 MIMO processing.}
\label{FIG:5}
\end{figure*}

\section{Single-Polarization Experimental Results}

We experimentally measured a 30-Gbaud QPSK signal after 60-km SMF transmission employing the proposed phase-retrieval receiver with noise loading.
Most parameters of the setup are the same as the ones used in the previous simulations.
A longer signal sequence of $2^{12}$ symbols is used.
Captured waveforms are oversampled at 60 GSamples/s, which gives two samples per symbol as a common oversampling ratio for pulse shaped signals.
30 million symbols are captured and used for post-processing and BER calculation.
Phase retrieval is applied in blocks of $2^{13}$ samples after 2 $\times$ oversampling and BER is calculated accordingly.
Analyzing longer sequences would require techniques such as the overlap-save method to enable block-wise and parallel processing of the data.

Histograms of the measured BERs under different OSNRs are shown in Fig.~\ref{FIG:4}(a)-(d).
The OSNR penalty at 2$\times$10$^{-2}$ is 12~dB and an error floor around 1.5$\times$10$^{-3}$ occurs as the OSNR greater than 25 dB.
This degradation compared to the simulations is primarily due to the electrical noise which can be expressed as an effective number of bits (ENOB), which is around 5.2 to 5.4 within the signal bandwidth.
To investigate the impact from the limited ENOB, Fig.~\ref{FIG:4}(e) compares the measured BERs (the mode of the histogram) with the simulated results assuming different ENOBs.
As indicated by the simulated results, an oscilloscope with a higher ENOB can dramatically improve the system performance.
Other imperfections such as incompletely compensated frequency roll-off and frequency response ripples from the components will also distort the waveforms and cause phase retrieval errors.
Dedicated component calibrations will be able to mitigate these imperfections.

\begin{figure}[!t]%1
    \centerline{\includegraphics[width=1.8in]{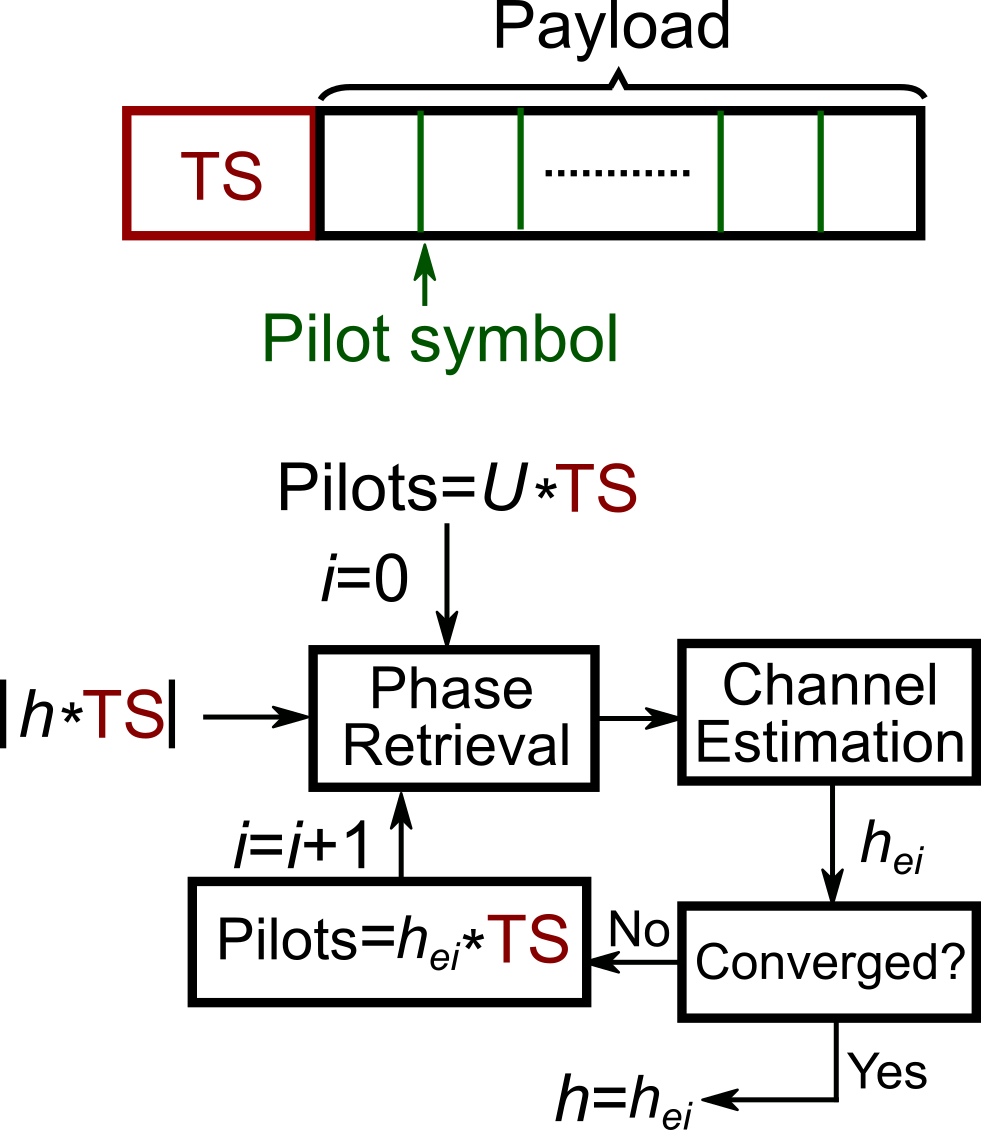}}
    \caption{~Illustration of the transmitted signal frame structure and 2$\times$2 time-domain transfer matrix $h$ estimation (TS: training sequence, *: convolution, $U$: unitary matrix and $h_{ei}$: estimated $h$ at $i_{th}$ iteration),
    }
    \label{FIG:6a}
\end{figure}

\begin{figure*}[!t]%1
\centerline{\includegraphics[width=6.8in]{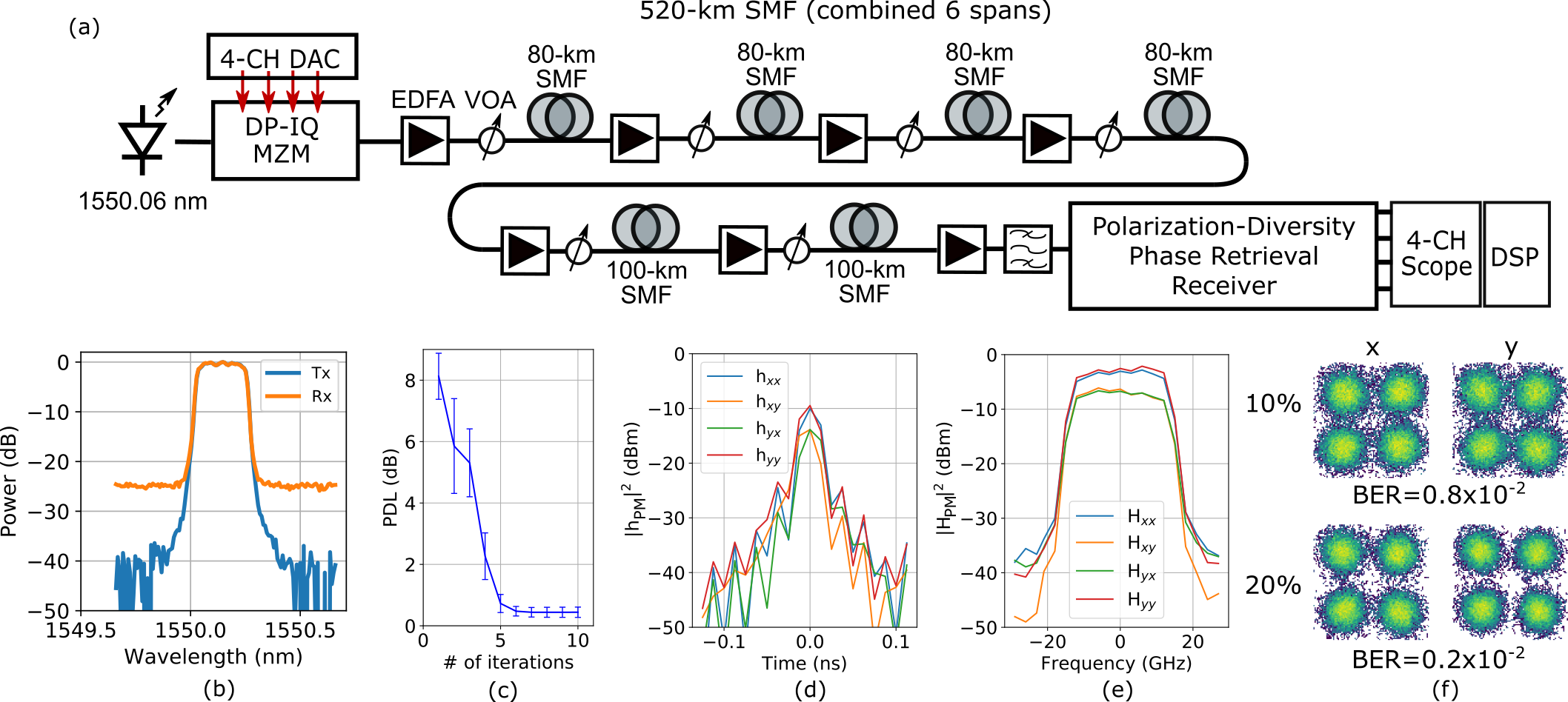}}
\caption{(a) Experimental setup for detecting a polarization-diversity QPSK signal after 520-km SMF transmission,
	(b) measured optical spectra at transmitter and receiver,
	(c) mean and standard deviation of the calculated PDL from 300 waveforms as a function of the number of iterations,
	(d) and (e) intensity of the estimated fiber channel matrix \textit{$h_{PM}$} and \textit{$H_{PM}$} in the time and frequency domain after removing chromatic dispersion,
	and (f) recovered QPSK constellations for both polarizations employing 10$\%$ (upper) and 20$\%$ (lower) pilot symbols.}
\label{FIG:6}
\end{figure*}

\section{Polarization-Diversity Phase-Retrieval Receiver}
The coherent receiver's ability to measure the amplitude and phase of both polarizations simultaneously enables polarization-division multiplexed transmission.
The polarization state at the output of the fiber is scrambled and can have strong wavelength dependence.
With the knowledge of the full fields, MIMO processing can unscramble the polarization and recover the polarization-division multiplexed launch signal.
This section experimentally demonstrates how the phase-retrieval receiver can accomplish this using intensity only measurements.
It takes two steps to perform phase retrieval for polarization-division multiplexed transmission: 1) 2$\times$2 fiber transfer matrix \textit{$h$} estimation employing the training sequence and 2) full-field recovery over the payload, followed by the traditional DSP including 2$\times$2 MIMO.

\subsection{Receiver Schematic}

Figure~\ref{FIG:5}(a) shows the schematic of the receiver architecture for detecting a polarization-diversity complex-valued signal using only intensity measurements.
A polarization beam splitter (PBS) separates the polarization mixed received signal onto the $x$ and $y$ polarizations with respect to the PBS axis.
The intensity of each polarization is measured before, $a_x(t) = |s_x(t)|^2$, and after, $b_x(t)=|d_x(t)|^2$ the dispersive elements.
These four real-valued signals, as illustrated in Fig.~\ref{FIG:5}(b) are passed to the DSP (Fig.~\ref{FIG:5}(c)), which is separated into two sections: a phase-retrieval front end to recover the full field of each polarization and a 2$\times$2 MIMO equalizer that can compensate for polarization mixing and polarization mode dispersion provided the phase is recovered properly.
% The pilot symbols are pre-calculated and fed back from the equalizer to assist the convergence of phase retrieval algorithm.
This receiver architecture can be scaled up to support space-division multiplexing~\cite{chen_mode-multiplexed_2019} by using a spatial multiplexer~\cite{huang_all-fiber_2015,chen_compact_2014,fontaine_laguerre-gaussian_2019} to demultiplex the spatial modes prior to polarization separation.

\subsection{Dual-Polarization Transmission Setup}

We transmitted a dual-polarization QPSK signal over a 520-km standard SMF span comprising of four 80-km and two 100-km SMF spans to highlight the phase-recovery receiver's ability to handle large amounts of CD and polarization mixing (PMD).
Figure~\ref{FIG:6}(a) shows the transmission experiment setup.
An external cavity laser with a linewidth of 100 kHz at 1550.06~nm was modulated by a dual-polarization IQ Mach-Zehnder modulator (DP-IQ MZM).
Nyquist-shaped polarization-diversity 30-Gbaud QPSK waveforms with a spectral roll-off of 0.1 and length of $2^{14}$ were generated by a 4-channel programmable digital-to-analog converter (DAC) operating at 60 GSamples/s.
The input power of each span is adjusted to 0~dBm via a variable optical attenuator (VOA) and the accumulated dispersion is 8921~ps/nm.
After the last span, the optical signal is amplified and spectrally filtered before being detected by the polarization-diversified phase-retrieval receiver (Fig.~\ref{FIG:5}(a)).
Two dispersion compensation fibers with a CD around 650 ps/nm are used as the dispersive elements.
After direct detection with four single-ended photo-diodes, the signals are sampled by a 4-channel digital sampling oscilloscope at 80~GSamples/s and undergo digital resampling afterwards.
Figure~\ref{FIG:6}(b) shows the measured optical spectra at the transmitter and receiver.
After 520-km SMF transmission, the OSNR at the receiver is around 28~dB.

\subsection{Fiber Coupling Matrix Estimation}

Due to the polarization coupling (PMD), a training sequence is used to reconstruct the fiber coupling matrix \textit{$h$} that contains polarization mixing and CD: \textit{$h$} = \textit{$h_{CD}*h_{PM}$}, where \textit{$h_{PM}$} is the matrix for polarization mixing after digitally removing \textit{$h_{CD}$} from \textit{$h$}.
\textit{$h$} is estimated by multiplying the received training symbols with the pseudo inverse of time-aligned transmitted training symbols~\cite{ryf_mode-division_2012}.
By running the phase retrieval algorithm, channel estimation, and pilot symbol replacement iteratively, \textit{$h$} can be properly recovered, as illustrated in Fig.~\ref{FIG:6a}.
% (See Supplement for details).
In order to verify the robustness and accuracy of the channel estimation, 300 training sequences (each with a length of $2^{14}$) are captured and used to estimate \textit{$h$}.
Polarization-dependent loss (PDL) is calculated based on the singular values of the estimated $h_{ei}$, which is the estimated $h$ at $i_{th}$ iteration.
A unitary matrix ($U$) is used as the initial condition which has zero PDL.
Figure~\ref{FIG:6}(c) shows the mean and standard deviation of the PDL as a function of the number of iterations.
Low PDL with a small deviation can be achieved after six iterations, which indicates that \textit{$h$} is properly estimated.
Figure~\ref{FIG:6}(d) and (e) shows the estimated polarization mixing matrix for one waveform after six iterations in time (\textit{$h_{PM}$}) and frequency (\textit{$H_{PM}$}) domain, respectively.
\textit{$h_{PM}$} is used to forward-propagate the pilot symbols, \textit{$\vec{p}$}, to the receiver where they can be applied to correct the payload.
Received training symbols containing fiber coupling information is recovered from the intensity measurements after employing phase retrieval.

\subsection{Experimental Results}
In the experiment, four intensity measurements as $d_x(t)$, $s_x(t)$, $d_y(t)$ and $s_y(t)$ are captured and resampled digitally to 3 million samples for each with a ratio of two samples per symbol.
The first $2^{14}$ symbols are applied as the training sequence and the remaining are treated as the payload.
Pilot symbols for retrieving the phases of the payload are updated using \textit{$h_{PM}$*$\vec{p}$} based on the estimated \textit{$h_{PM}$} after applying channel estimation over the training sequence with six iterations.
We use the overlap-save method with a 50$\%$ save to enable block-wise phase retrieval with a block length of $2^{10}$.
After reconstructing the full fields of the polarization-diversity signal using phase retrieval, CD from the transmission fiber is first compensated prior to the 2$\times$2 frequency-domain equalizer with symbol-spaced taps.
Data-aided least-mean-square (LMS) algorithm is used initially to force the convergence of the equalizer, followed by constant-modulus algorithm (CMA)~\cite{benvenuto_algorithms_2002}.
Carrier-phase recovery and bit-error-rate (BER) counting are performed afterwards.
Figure~\ref{FIG:6}(d) shows the recovered QPSK constellations for both polarizations employing 10$\%$ (upper) and 20$\%$ (lower) pilot symbols, respectively.
After 520-km transmission, the BER for employing 10\% and 20\% pilot symbols was 0.8$\times 10^{-2}$ and 2.2$\times 10^{-3}$.
Assuming a 20$\%$ overhead for forward error correction (FEC) and 10$\%$ for phase retrieval, a net data rate of 84~Gbit/s is achieved.
System performance is mainly limited by the electrical noise from the oscilloscope (ENOB) and lower than optimum photo-currents due to the losses from the optical splitters and dispersive elements within the receiver.

\section{Conclusion and Outlook}
We present and experimentally demonstrate an optical receiver for coherent communications that can reconstruct complex-valued signals using only direct detection and without the use of any optical carriers.
Signal phases are recovered based on alternating projection employing dispersion.
The proposed phase-retrieval receiver only requires $2N$ single-ended photo-diodes to detect the intensities of $N$ scrambled spatial and polarization modes but can perform as conventional optical coherent receivers to accomplish linear impairments compensation and support polarization- and space-division multiplexing.

One attractive feature of employing dispersion for phase retrieval is that the required amount of dispersion 
reduces quadratically as a function of the transmitted symbol rate, and the current industry trends in optical
communication is towards higher baudrates~\cite{yamazaki_digital-preprocessed_2016}.
State of the art optical transmitters offer a signal baud rate as fast as 180~Gbaud~\cite{raybon_180-gbaud_2018}, which can lower the required dispersion significantly to 288~ps/nm to achieve $<$ 4-dB penalty or 18~ps/nm for a performance comparable to the presented experiment.

The phase recovery algorithm compares two intensity measurements of the same complex-valued signal with and without experiencing a known projection.
With the aid of pilot symbols, accurate phase can be reconstructed as both intensity waveforms are detected without any distortions.
Either additive white Gaussian noise from optical amplification or quantization noise due to the limited ENOB during analog-to-digital or digital-to-analog conversion introduces waveform distortions, which will weaken the pilot symbol constraint and cause additional OSNR penalties.
Another source of waveform distortion is the fiber nonlinearity, especially after long-distance transmission.
Further work will integrate nonlinear impairment mitigation into the phase retrieval algorithm.
The dispersion compensating fibers used in the experiments have a small effective area and therefore nonlinear impairments can be easily introduced if a large input signal power is used.
Alternative dispersive element solutions could be on-chip optical all-pass filters~\cite{madsen_multistage_1999} or Bragg gratings~\cite{giuntoni_continuously_2012}, which provide sufficient amount of dispersion and permit future photonic integration of the proposed phase-retrieval receiver.

Latency will be introduced in the system since the phase retrieval algorithm finds the optimum phase in an iterative manner.
The convergence speed of the algorithm can be enhanced through optimizing the parameters such as the acceptable error level $\epsilon$ and the iteration number between the two phase resets.
Introducing more constraints using the signal properties such as its constellation pattern can also help to reduce the required number of iterations effectively. 
It is beneficial in improving the convergence speed, phase errors and receiver performance to explore parallel alternative projections employing multiple dispersive elements~\cite{brady_nonlinear_2006,rodrigo_multi-stage_2010} with a cost of higher hardware complexity, see Appendix. D for a detailed description.
It would also be worthwhile exploring the combination of the phase retrieval algorithm with convex optimization~\cite{hayakawa_reconstruction_2018} or machine learning~\cite{rivenson_phase_2018} and further optimize the interaction between the phase-retrieval front-end and the following DSP blocks.

\begin{figure*}[!t]%1
\centerline{\includegraphics[width=6.5in]{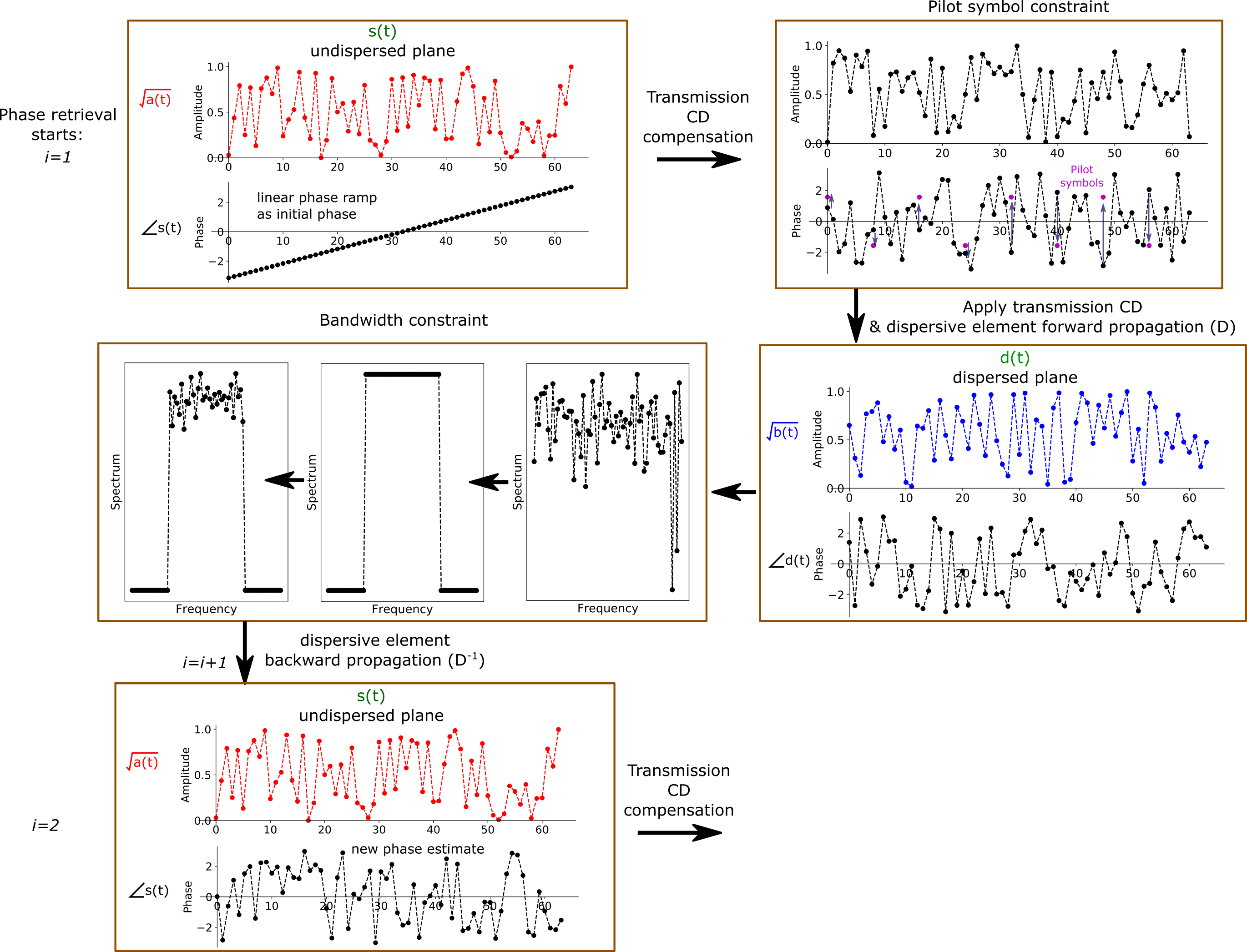}}
\caption{~Illustration of spectral and pilot symbol constraints in phase retrieval algorithm.
}
\label{FIG:1b}
\end{figure*}

\begin{figure}[!b]%1
\centerline{\includegraphics[width=3.0in]{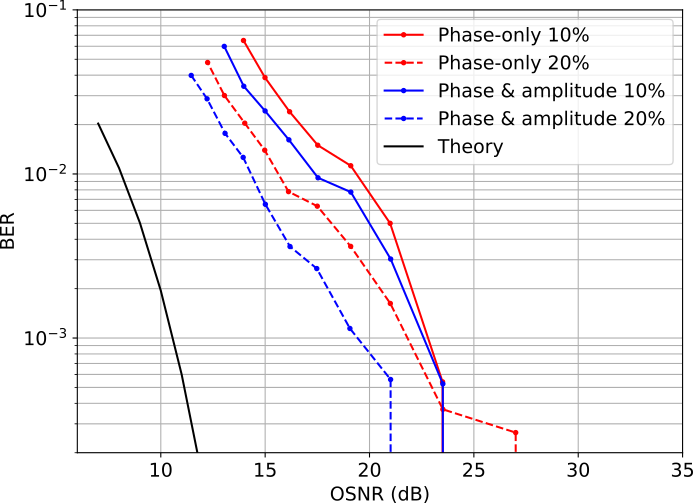}}
\caption{~Simulated BER versus OSNR using only the phase or both phase and amplitude (full field) information of the 10$\%$ or 20$\%$ pilot symbols.
}
\label{FIG:SA}
\end{figure}

\begin{figure}[!b]%1
\centerline{\includegraphics[width=3.0in]{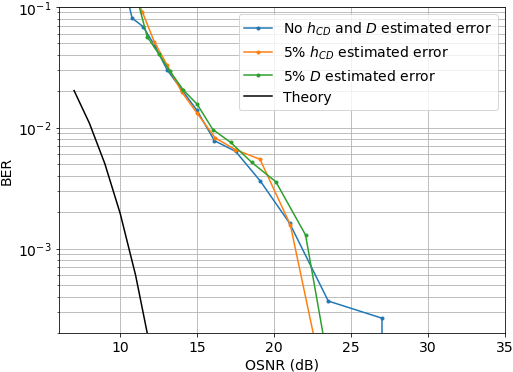}}
\caption{~Simulated BER versus OSNR for detecting a single-polarization 30-Gbaud QPSK signal with \textit{$h_{CD}$} and dispersive element $D$ estimation errors.
}
\label{FIG:SB}
\end{figure}

\begin{figure*}[!t]%1
\centerline{\includegraphics[width=6.5in]{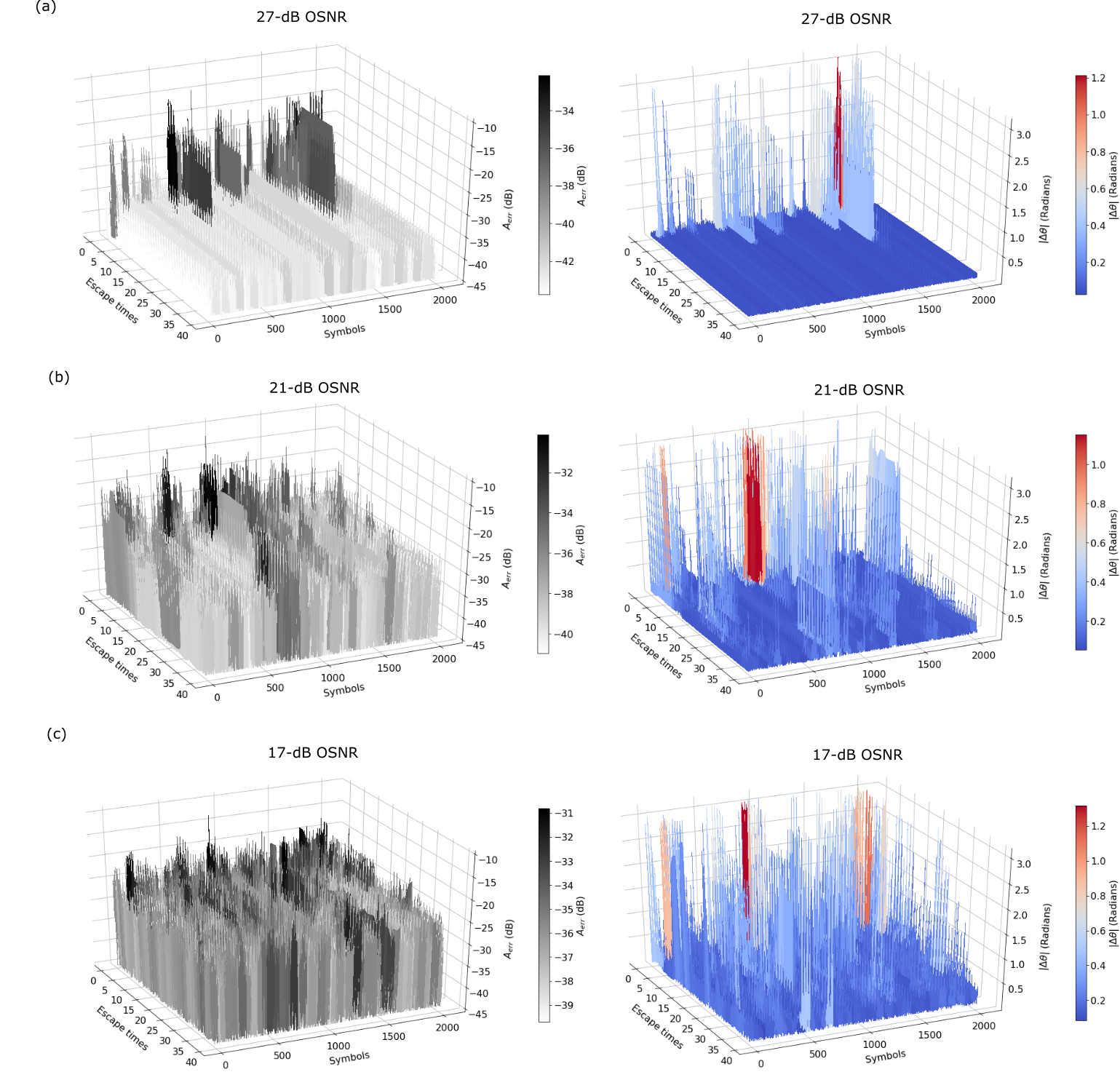}}
\caption{
$A_{err}$ (left) and absolute phase error $|\Delta\theta|$ (right) for all the symbols after each escape with an OSNR of (a) 27~dB, (b) 21~dB and (c) 17~dB.
}
\label{FIG:S2}
\end{figure*}

\begin{figure}[!t]%1
\centerline{\includegraphics[width=3.0in]{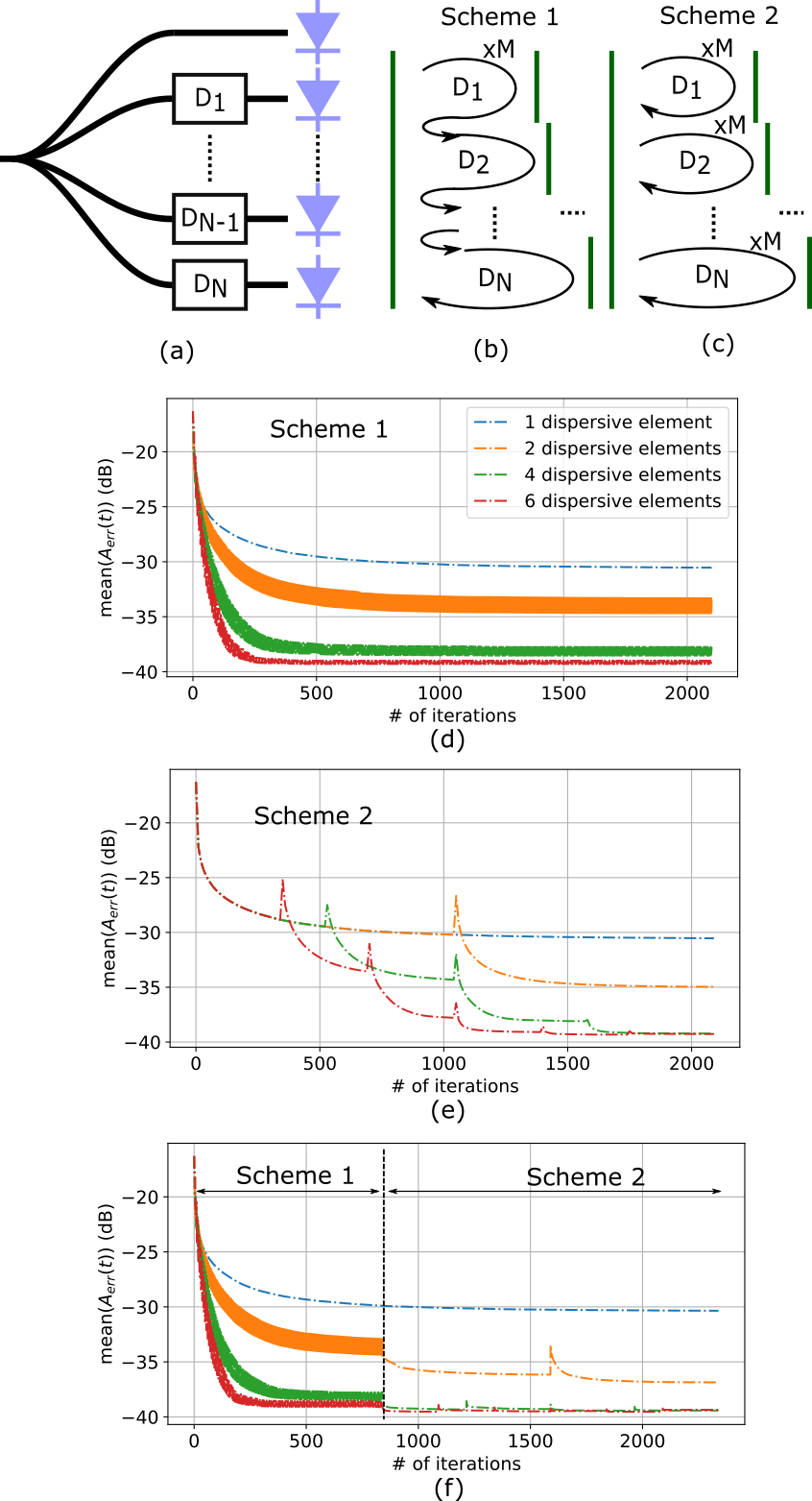}}
\caption{~(a)~Illustration of a phase-retrieval receiver employing parallel alternative projections, (b) and (c) schematic of two schemes running the modified GS algorithm iteratively between multiple dispersive elements,
(d-f)~mean $A_{err}(t)$ averaged over $2^{11}$ symbols in dB versus the number of iterations at 30-dB OSNR using scheme 1, 2 and combined.
}
\label{FIG:SD}
\end{figure}

\section{Appendix}

\subsection{Spectral and Pilot Symbol Constraints}
Figure~\ref{FIG:1b} explains the application of the spectral and pilot symbol constraints in the phase retrieval algorithm.
The example uses 64 samples for illustration.
Square root of the directly measured $a(t)$ is used as the amplitude of $s(t)$.
For the sake of illustration, a linear phase ramp is assumed as the initial phase condition.
After compensating the transmission CD \textit{$h_{CD}$} of $s(t)$, pilot symbols are applied to update the phases of the samples at the pre-determined locations.
In this case, every $8^{th}$ symbol's phase is replaced with a pilot symbol’s phase.
The amplitude of the pilot symbol can still be used to encode information.
System performance could be further improved as both phase and amplitude of the pilot symbol are applied as the constraints in phase retrieval, see Fig.~\ref{FIG:SA}.
The simulations are for detecting a single-polarization 30-Gbaud QPSK signal transmitted over 60-km SMF.
% and start from zero.
The amplitude of $d(t)$ at the dispersed plane is replaced by the measured $b(t)$ after forward propagation over the transmission fiber and the dispersive element ($D$).
With the knowledge of the transmitted signals' bandwidth and shaping scheme, spectral filtering is applied as the bandwidth constraint.
\textit{$h_{CD}$} compensation, and spectral filtering are all accomplished in the frequency domain using fast Fourier transform and its inverse.
The algorithm repeats itself until it reaches the maximum iteration number.

\subsection{\textit{$h_{CD}$} Estimation}

\textit{$h_{CD}$} can be modeled as a transfer function with a quadratic phase response and constant amplitude, written as: \textit{$h_{CD}=exp(-jB_2t^2/2)$}, where $B_2$ is the first-order dispersion coefficient.
\textit{$h_{CD}$} can be characterized in advance~\cite{costa_phase_1982,ponzo_fast_2014} or measured directly using the phase-retrieval receiver through intensity correlation.
The value of $B_2$ from transmission can be found as the correlation between $a(t)$ and $|\textit{$h_{CD}$}*x(t)|^2$ reaches maximum, where $x(t)$ is the transmitted signal.
To characterize the impact from \textit{$h_{CD}$} and dispersive element $D$ estimation inaccuracies on system performance, we simulated BERs versus OSNR for detecting a single-polarization 30-Gbaud QPSK signal transmitted over 60-km SMF using 20$\%$ pilot  symbols.
It can be observed in Fig.~\ref{FIG:SB} that the phase-retrieval receiver is resilient to small \textit{$h_{CD}$} and $D$ estimation errors.

\subsection{Symbol-Wise Phase Retrieval Error versus OSNR}
Phase retrieval error for each symbol is calculated by $A_{err}(t)= |a(t)-|s'(t)|^2|^2$.
At a low OSNR region, signal is polluted with more noise, which can be treated as waveform distortions in phase retrieval so that the recovered phase accuracy is reduced.
Therefore, the phase retrieval error increases as the OSNR decreases.
Figure~\ref{FIG:S2}(a)-(c) shows $A_{err}$ (left) and absolute phase error $|\Delta\theta|$ (right) for all the symbols after each escape up to 40 for an OSNR of 27~dB, 21~dB and 17~dB, respectively.

\subsection{Parallel Alternative Projections}
Parallel alternative projections use more dispersive elements, as illustrated in Fig.~\ref{FIG:SD}(a).
Running the modified GS algorithm iteratively between the multiple dispersive elements provides more intensity constraints, which enables fast algorithm convergence and avoids local minima.
Figure~\ref{FIG:SD}(b) and (c) provides two iteration schemes.
Both use the retrieved phase from the current iteration as the start phase condition for a new iteration but have different iteration paths.
% but experience different loop paths.
% Each iteration in the $1^{st}$ scheme has a large loop covering all the projections from $D_1$ to $D_N$.
% The $2^{nd}$ scheme runs $M$ times iterations over one projection then uses the measured intensity from another projection.
We carried out simulations for detecting a single-polarization 30-Gbaud QPSK signal transmitted over 60-km SMF using 20$\%$ pilot symbols.
Phase reset is not applied.
The amount of dispersion of $D_n$ is $n\Delta D$, where $n$ is the order of the dispersive elements from 1 to $N$ and $\Delta D$ is the dispersion increment, which is 325~ps/nm, larger enough to introduce symbol mixing and intensity variations at different dispersed planes.
Figure~\ref{FIG:SD}(d) and (e) show the mean $A_{err}(t)$ averaged over $2^{11}$ symbols and 100 noise realizations in dB versus the number of iterations at 30-dB OSNR for both schemes with a total number of iterations ($NM$) of 2100.
For example, the algorithm will run 1050 iterations ($M$=1050) as two dispersive elements are applied ($N$=2).
In the $2^{nd}$ scheme, after looping over one projection $M$ times, a sub-optimal solution can be reached, which introduces transition error spikes as moving to a new projection, see Fig.~\ref{FIG:SD}(e).
Scheme 1 has a faster convergence rate but is numerically less stable compared to the $2^{nd}$ scheme.
Combining both schemes sequentially can help to find the optimum solution with a reduced number of iterations, see Fig.~\ref{FIG:SD}(f).

\def\url#1{}

\bibliographystyle{IEEEtran_JLT}

\bibliography{IEEEabrv,JLT_PhaseRetrieval}

\end{document}